\documentstyle[11pt,epsf]{article}

\advance \topmargin by -\headheight
\advance \topmargin by -\headsep

\evensidemargin \oddsidemargin
\newfont{\headfont}{cmbx10 scaled 1440}
\newfont{\namefont}{cmr10}
\newfont{\initialfont}{cmr10 scaled 1200}
\newfont{\addfont}{cmti7 scaled 1440}
\newfont{\boldmathfont}{cmbx10}
\newfont{\figfont}{cmr7 scaled 1200}

\let\nopictures=Y
\newfont{\headfontb}{cmbx10 scaled 1728}
\begin{document}
\begin{titlepage}
\renewcommand{\thefootnote}{\fnsymbol{footnote}}
\begin{center}
{\headfontb Time-dependent quantum scattering\\ ~in 2+1 dimensional gravity}
\footnote{This work is supported in part by funds provided by the
U. S. Department of Energy (D.O.E.) under cooperative agreement
\#DE-FC02-94ER40818.}

\end{center}
\vskip 0.3truein
\begin{center}
{
{\Large M.}{} {\Large A}{lvarez\footnote{Email: marcos@mitlns.mit.edu}}
{, }
{\Large F.M.}{ de} {\Large C}{arvalho} {\Large F}{ilho
\footnote{Email: farnezio@mitlns.mit.edu. Permanent address: Escola Federal de
Engenharia de Itajub\'a, C.P.50, Itajub\'a-M.G., Brazil}}
{, and }
{\Large L.}{} {\Large G}{riguolo\footnote{Email: griguolo@mitlns.mit.edu}}
}
\end{center}
\begin{center}
{\addfont{Center for Theoretical Physics}}\\
{\addfont{Massachusetts Institute of Technology}}\\
{\addfont{Cambridge, Massachusetts 02139 U.S.A.}}
\end{center}
\vskip 0.5truein
\begin{center}


\bf ABSTRACT
\end{center}

The propagation of a localized wave packet in the conical space-time created
by a pointlike massive source in 2+1 dimensional gravity is analyzed. The
scattering amplitude is determined and shown to be finite along the classical
scattering directions due to interference between the scattered and the
transmitted wave functions. The analogy with diffraction theory is emphasized.

\vskip 7truecm
\leftline{CTP \# 2451  \hfill July 1995}
\smallskip
\leftline{hep-th/9507134}

\end{titlepage}
\setcounter{footnote}{0}


\newpage

\section{Introduction}

\indent The time-dependent scattering problem was solved in the case of the
Aharonov-Bohm interaction in \cite{steli}. This author considered the time
evolution of an electrically charged well-localized wave packet in presence of
a magnetic vortex.  The main result in that work
is the analysis of the forward direction, where the wave packet undergoes
a self-interference; the probability density current was shown to be finite.

The question arises if a similar analysis can be carried out in 2+1
dimensional gravity. By this we mean to consider the scattering of a wave
packet by a static source in planar gravity, to find the scattering
amplitude, and to determine the behaviour of the wave packet along the
directions where self-interference effects are significant.

The classical theory of 2+1 dimensional gravity, as well as its
interpretation as a conical space-time, was presented
in \cite{dejath}. The quantum-mechanical scattering problem for two scalar
particles interacting only gravitationally in 2+1 dimensions was first solved
in \cite{thooft} by reducing the problem to the motion of a free
particle on a cone. A closely related procedure was put forward in
\cite{desjac}, this time derived from a partial wave decomposition. Needless
to say, both methods yield the same scattering amplitude. These works showed
that in the case of 2+1 dimensional gravity the forward direction is
not exceptional; it is at the classical scattering angles where
self-interferences take place.

A further step was taken in \cite{gerjac}. These authors not only generalized
the previous results to the case of spinning sources, but also pointed out an
interesting analogy between scattering in 2+1 dimensional gravity and
classical diffraction theory. Even though their discussion of this point is
qualitative, they were able to interpret the main features of the scattering
amplitude as a diffractive effect.

Albeit these works provided a thorough understanding of the
scattering process, none of them addresses the time-dependent scattering
problem as posed before. In this work we present a solution based in the
optical analogy noted in \cite{gerjac}.

This paper is organized as follows. In Sect.~2 we recall the propagator found
in \cite{desjac} for the conical Schr\"odinger equation, and analyze its
behaviour close to the classical scattering angles. This is accomplished by
means of a method developed by W. Pauli in the context of classical
diffraction theory \cite{pauli}. In Sect.~3 we introduce an incoming
Gaussian wave packet, with vanishing impact parameter, and study its
propagation by using the results of the previous Section.  The
result is free from the singularities in the scattering
amplitude found in \cite{thooft}, \cite{desjac}. We find a
cancellation of finite
discontinuities along the classical scattering angles due to interference
between scattered and transmitted waves. This can be considered a
quantitative version of the qualitative analysis
presented in \cite{gerjac}. In Sect.~4 we perform a similar
analysis for a wave packet
with non-zero impact parameter. Finally, in Sect.~5 we present our
conclusions. In the Appendix the same method is applied to time-independent
scattering.

\section{Calculation of the propagator}
\indent In this Section we shall discuss the quantal propagator for a
test-particle of mass $m$ moving in the conical space created by a static
mass $M$ at the origin of our coordinate system. We refer the reader to
\cite{dejath} and \cite{desjac} for a full exposition of these points.

Let us summarize the geometrical structure of the space-time
in question. An intrinsic characterization \cite{dejath}
uses a Euclidean metric with incomplete angular range to describe the
two-dimensional geometry of space:
\begin{equation}
(dl)^{2}=(dr)^{2}+r^2(d\varphi)^2,
\qquad\qquad -\pi\alpha\leq\varphi\leq \pi\alpha,
\label{21}
\end{equation}
where $0\leq(1-\alpha)=4MG<1$ and $G$ is ``Newton's constant''. We
recall that in this situation (quantal
scattering of a test-particle by a static mass) the time-component of the
metric does not play any role. An alternative characterization of this conical
space is based on embedded coordinates \cite{desjac}
\begin{equation}
(dl)^{2}=\alpha^{-2}(dr)^{2}+r^2(d\theta)^2,
\qquad\qquad -\pi\leq\theta\leq\pi,
\label{22}
\end{equation}
We shall use these coordinates in the following because the full angular range
allows for conventional partial-wave analysis and identification of phase
shifts in the wave functions. The Hamiltonian of a test particle of mass $m$
in this conical space-time is
\begin{equation}
H=-{\hbar^2\over 2m}\Bigl[\alpha^2{1\over r}\,\partial_r(r\partial_r)+{1\over
r^2}\,\partial^2_\theta\Bigr].
\label{23}
\end{equation}
This operator is diagonalized by eigenfunctions proportional to Bessel
functions; the dependence in the angle $\theta$ factorizes in a
single-valued exponential:
\begin{eqnarray}
\Psi_{n,k}(r,\theta)&=&\sqrt{{1\over 2\pi}}\,\,
e^{i n\theta}\,u_n(kr)\nonumber\\
u_n(kr)&=&(-1)^{{n-|n|\over2}}J_{{|n|\over\alpha}}(kr),
\label{24}
\end{eqnarray}
where $k=2mE/ \hbar^2\alpha^2$, $E$ is the energy eigenvalue, and $n$ is an
integer. The radial
eigenfunctions $u_n(kr)$ are regular at the origin and have the following
asymptotic behaviour:
\begin{equation}
u_n(kr)\stackrel{kr\rightarrow\infty}{\longrightarrow}
\sqrt{{2\over \pi k r}}\cos\Big(kr-{|n|\pi\over 2\alpha}-{\pi\over 4}
+{(|n|-n)\pi\over 2}\Big).
\label{asym}
\end{equation}
Thus the phase shifts are independent of the energy of the incoming particle,
as a consequence of non-relativistic conformal invariance,
and increase with $|n|$,
\begin{equation}
\delta_n =-{|n|\pi\over 2}(\alpha^{-1}-1).
\label{phase}
\end{equation}

Since we are interested in a time-evolution problem, we need the Feynman
propagator
\begin{equation}
G({\bf r},{\bf r}';t)=<{\bf r}'|e^{-i Ht}|{\bf r}>,
\label{25}
\end{equation}
having a spatial delta function as boundary condition at $t=0$. Using the
complete set of energy eigenstates and taking as initial and final points
${\bf{r}}=(r,\theta)$ and ${\bf{r'}}=(r', \theta')$,
we have the representation (the imaginary
time $T=i t$ makes well-defined the integration)
\begin{equation}
G({\bf r},{\bf r}';-i T)={1\over 2\pi}\int_{0}^{\infty}k\,dk\,
e^{-{\hbar\alpha^2 T k^2\over 2m}}\sum_n J_{{|n|\over\alpha}}(kr)\,
J_{{|n|\over\alpha}}(kr')\,e^{i n(\theta'-\theta)}.
\label{26}
\end{equation}
The integration leads to the Deser-Jackiw propagator. Going back to real
time $t$ this propagator can be written as
\begin{equation}
G(r,\theta;r',\theta';t)=
{m\over{2\pi i \hbar t\alpha^2}}\exp\Big{\{}{i m\over
{2\hbar t\alpha^2}}(r^2+r'^2)\Big{\}}\sum_n e^{i n(\theta'-\theta)}
I_{|n|\over \alpha}
\Big({mrr'\over i \hbar t\alpha^2}\Big).
\label{ge}
\end{equation}
The partial wave sum can be evaluated with the help of the Schl\"afli
contour integral representation for the Bessel function, whose
contour of integration is shown in Fig.~1,
\begin{equation}
I_{\nu}(x)={1\over 2\pi}\int_C dz \,e^{x\cos z +i \nu z}.
\label{schl}
\end{equation}

\begin{figure}[th]
\centerline{\hskip.4in\epsffile{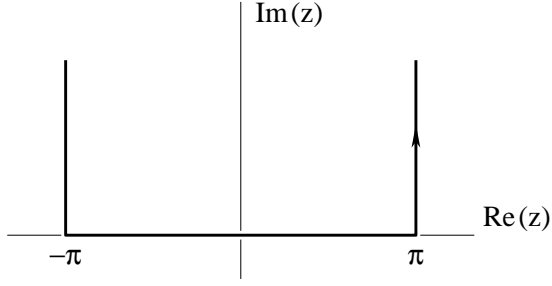}}
\caption{The Schl\"afli contour}
\end{figure}

After the summation the propagator $G(r,\theta;r',\theta';t)$ can be written
as a sum of two different terms, namely $G_1$ and $G_2$, corresponding
respectively to the transmitted and the scattered wave:
\begin{eqnarray}
\lefteqn{G_1(r,\theta;r',\theta';t)=} \nonumber \\
& &{m\over 2\pi i \hbar t\alpha}\mathop{{\sum}'}_n \exp\Big{\{}i
{m\over 2\hbar t\alpha^2}\big[ r^2+r'^2-2rr'\cos \alpha(\theta'-\theta-2\pi n)
\big]\Big{\}}, \nonumber \\
\lefteqn{G_2(r,\theta;r',\theta';t)=} \nonumber \\
& &{m\over 8\pi^2 i \hbar
t\alpha^2}\int\limits_{-\infty}^{\infty} dy\Big{\{}\cot\Big[
{i y\over 2\alpha}-
{\pi\over 2\alpha}+{\theta'-\theta\over 2}\Big]- \nonumber \\
& &-\cot\Big[ {i y\over 2\alpha}+
{\pi\over 2\alpha}+{\theta'-\theta\over 2}\Big]\Big{\}}\exp\Big{\{}i
{m\over 2\hbar t\alpha^2}(r^2+r'^2+2rr'\cosh y)\Big{\}}, \label{geundos}
\end{eqnarray}
where the primed sum includes only $n$
such that $\alpha(\theta'-\theta-2\pi n)
\in (-\pi, \pi).$ The propagator $G_1$ is presented in a closed form, but
$G_2$ is given as an integral representation. We are going to elaborate the
latter in order to make it useful for calculations. The propagator
$G_2$ can be written in an alternative way by means of a trigonometric
identity:
\begin{eqnarray}
G_2(r,\theta;r',\theta';t)&=&{m\over 4\pi^2 i \hbar
t\alpha^2}\int\limits_{-\infty}^{\infty} dy{\sin{\pi\over\alpha}\over
\cos{\pi\over\alpha}-\cos\Big({i y\over\alpha}+\theta'-\theta\Big)} \nonumber
\\
& &\times\exp\Big{\{}i
{m\over 2\hbar t\alpha^2}(r^2+r'^2+2rr'\cosh y)\Big{\}}.
\label{gedosdos}
\end{eqnarray}
Therefore, if $\alpha^{-1}$ is an integer this contribution to the propagator
vanishes. Otherwise the integral can be performed in the
limit of large $mrr'/\hbar t$, where the leading contribution
comes from the small-$y$ region:
\begin{eqnarray}
G_2(r,\theta;r',\theta';t)&\approx&{m\sin{\pi\over\alpha}\over 4\pi^2 i \hbar
t\alpha^2}\exp\Big{\{}{i m\over 2\hbar t\alpha^2}(r^2+r'^2)\Big{\}}
\int\limits_{-\infty}
^{\infty}dy\,\,\exp\Big{\{}{i mrr'\over 2\hbar t\alpha^2}
y^2\Big{\}} \nonumber \\
& &\times{1\over\cos{\pi\over\alpha}-\cos(\theta'-\theta)+{i y\over\alpha}
\sin(\theta'-\theta)+
{\cal{O}}(y^2)}.
\label{gedostres}
\end{eqnarray}

To proceed with the integration we need an additional assumption. We shall
consider the case $\theta'-\theta \neq \pm \pi/\alpha\,\, {\rm{mod(2\pi)}}$,
{\it{i.e.}}, we keep away from the classical scattering
angles \cite{desjac}. This allows to approximate the integral by a
Gaussian. The final result is
\begin{equation}
G_2(r,\theta;r',\theta';t)\approx \Big({m\over8\pi^2\hbar t\alpha^2
i rr'}\Big)^{1\over2}{\sin{\pi\over\alpha}\over
\cos{\pi\over\alpha}-\cos(\theta'-\theta)}\exp\Big{\{}{i m\over 2\hbar
t\alpha^2}(r+r')^2\Big{\}},
\label{gedosfin}
\end{equation}
which can be used immediately to find the scattering amplitude. The result is
that of \cite{thooft} and \cite{desjac}.

Now we analyze Eq.~(\ref{gedosdos}) in the
vicinity of the classical scattering angle. A straightforward saddle-point
calculation is not possible now because the integrand develops a singularity
precisely at the saddle-point $y=0$. Hence the problem arises to obtain an
explicit formula for the scattered propagator $G_2$ in the limit of large
$mrr'/\hbar t$, which will be valid also at the classical scattering
directions. It is here that the method
developed in \cite{pauli} comes into play. Pauli considered the problem of the
diffraction of light by a wedge limited by two perfectly reflecting
planes. The
diffracted wave can be calculated by means of an integral representation
similar to Eq.~(\ref{gedosdos}), whose singularity lies in the
boundary between
the ``illuminated'' region and the ``shadow'' of geometrical optics. He
was able to show that the transition from shadow to light is
completely smooth. Our problem is to show
that the apparent singularity present in Eq.~(\ref{gedosfin}) when
$\theta'-\theta =
\pm \pi/\alpha\,\, {\rm{mod(2\pi)}}$ does not actually exist, so
that the wave function is regular everywhere. The formal similarity between
these two problems makes possible to apply Pauli's method in our case.

\begin{figure}[th]
\centerline{\hskip.5in\epsffile{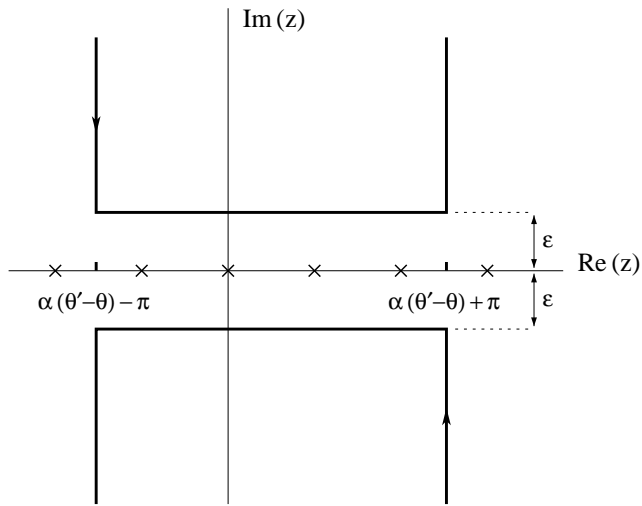}}
\caption{The contour for the propagator}
\end{figure}

\begin{figure}[th]
\centerline{\hskip.5in\epsffile{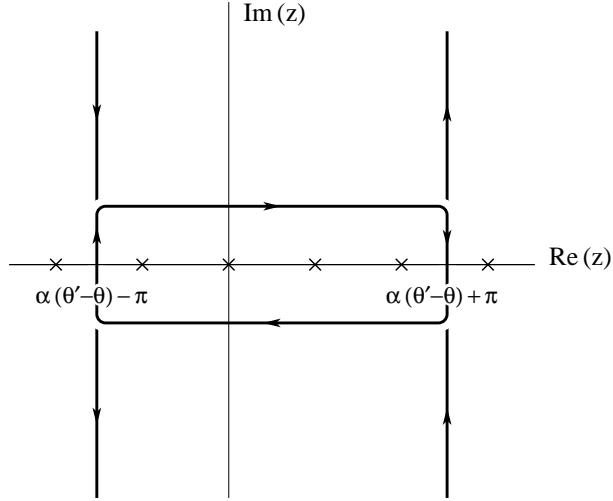}}
\caption{An equivalent contour}
\end{figure}

Let us first examine the solution given in \cite{desjac} to a similar
difficulty in the time-independent scattering of plane waves in 2+1
dimensional gravity. These authors started with an integral
representation for the wave function whose integration path is that of
Fig.~2 or, equivalently, that of Fig.~3 (see \cite{desjac} for
details). The equivalence between these contours follows from the
cancellation of the vertical sides of the closed contour in Fig.~3 with the
adjacent segments of the straight lines. All the singularities of the
integrand are poles which lie on the real axis at $z=2\pi\alpha N$, being $N$
an integer. The closed contour in
Fig.~3 corresponds to a sum of Cauchy residues, which yields the
transmitted wave; the two straight lines correspond to the scattered
wave.

This construction is rigurous as long as the contours
can be deformed to avoid the singularities. If $\theta'-\theta =
\pm \pi/\alpha\,\, {\rm{mod(2\pi)}}$ the contours cross over one of
the poles, which therefore cannot be avoided. In other words,
the decomposition of the wave function in ``transmitted'' (closed contour in
Fig.~3) and  ``scattered'' (straight lines, {\it{ibid.}}) components
must be re-examined at the
classical scattering angles. In \cite{desjac} it is assumed that the pole that
is now present at the boundary of the closed contour contributes only half its
residue, and that the two straight lines in Fig.~3 exclude
${\rm{Im}}(y)\in[-\epsilon, \epsilon]$, thereby not cancelling the vertical
sides of the closed contour. If we apply this idea, the integration
in Eq.~(\ref{geundos}) would be interpreted as a principal value.

Under this assumption the angular dependence
in Eq.~(\ref{gedosfin}) reduces to $-\cot (\pi/\alpha)$. Nevertheless, the
contours in Figs. 2 and 3 cannot be identified if the vertical sides of the
closed contour in Fig.~3 remain uncancelled. We conclude that this
analysis of the dominant (at large $t$) portion of the propagator close to the
classical scattering directions is not adequate for analyzing the physics: a
subdominant term in the scattered wave, found below, is essential.

In order to apply the method proposed in \cite{pauli} we go back to
Eq.~(\ref{gedosdos}) and change to a new set of variables
\begin{eqnarray}
y&=&i \eta \nonumber \\
{mrr'\over \hbar t\alpha^2}&=& \rho \nonumber\\
\theta'-\theta &=&-{\phi\over\alpha}
\label{change}
\end{eqnarray}
which gives an integral representation for $G_2$ more suitable for the
following analysis:
\begin{eqnarray}
G_2(r,\theta;r',\theta';t)&=&-{m\sin{\pi\over\alpha}\over 4\pi^2 \hbar
t\alpha^2}
\exp\Big{\{}i
{m\over 2\hbar t\alpha^2}(r^2+r'^2)\Big{\}}  \nonumber\\
& &\times\int\limits_{-i \infty+\gamma}^{i \infty-\gamma}
d\eta{e^{i \rho\cos\eta}
\over
\cos{\pi\over\alpha}-\cos\Big({\phi+\eta\over\alpha}\Big)},
\label{gedosgo}
\end{eqnarray}
where $\gamma$ is any angle between zero and $\pi$. The physically interesting
case is that of large $\rho$, where the method of steepest descent can be
applied. This requires the introduction of the variable
\begin{equation}
s=e^{i \pi/4}2^{1\over 2}\sin{\eta\over 2}.
\label{ese}
\end{equation}
As a path of integration, the real $s$ axis can be taken, so that $G_2$
becomes
\begin{eqnarray}
G_2(r,\theta;r',\theta';t)&=&{m\sin{\pi\over\alpha}\over 4\pi^2 \hbar
t\alpha^2}
\exp\Big{\{}i
{m\over 2\hbar t\alpha^2}(r^2+r'^2)\Big{\}}e^{-i\pi/4}2^{1\over2} \nonumber\\
& &\times\int\limits_{-\infty}^{\infty} ds{e^{i\rho}e^{-\rho s^2}
\over
\cos{\pi\over\alpha}-\cos\Big({\phi+\eta\over\alpha}\Big)}
\Big(1+{i\over2}s^2\Big)^{-{1\over2}}.
\label{gedosgogo}
\end{eqnarray}
The purpose of the preceding changes of variable was to extract the
Gaussian factor $\exp(-\rho s^2)$ now present in Eq.~(\ref{gedosgogo}). The
obvious procedure would be to expand the integrand, except the Gaussian
factor, in powers of $s$ and evaluate the integrals. The result obtained in
this way would become ill-defined
if $\theta'-\theta=\pm\pi/\alpha \,\, {\rm{mod(2\pi)}}$, which
corresponds to the classical scattering directions.

The method presented in \cite{pauli} avoids this difficulty by developing
not the whole integrand, but only a factor regular at the saddle point. If
we introduce the notation $-a=1+\cos\phi$, the propagator $G_2$ can be written
as
\begin{eqnarray}
G_2(r,\theta;r',\theta';t)&=&-{m\sin{\pi\over\alpha}\over 4\pi^2 \hbar
t\alpha^2}
\exp\Big{\{}i
{m\over 2\hbar t\alpha^2}(r^2+r'^2)\Big{\}} \nonumber\\
& &\times e^{i(\rho+\pi/4)}2^{1\over2}
\int\limits_{-\infty}^{\infty} ds\,\,e^{-\rho s^2}
{f(s,\phi)\over
i a+s^2},
\label{gedosult}
\end{eqnarray}
where the function $f(s,\phi)$ is defined as
\begin{equation}
f(s,\phi)={\cos\eta(s)+\cos\phi\over
\cos{\pi\over\alpha}-\cos{\phi+\eta(s)\over\alpha}}\,\,{1\over\cos{\eta(s)
\over 2}}.
\label{efe}
\end{equation}
This function is regular at the saddle point $\eta(s)=0$ even
if $\phi=\pm\pi$. Its only singularities at $\eta(s)=0$ occur
if $\phi=\pi+2\pi\alpha N$ with
$N$ integer but $\alpha N$ not integer. Nevertheless these cases will not be
relevant in our problem, since we are mainly interested
in $\phi\approx\pm\pi$.

Let us expand $f(s,\phi)$ in powers of $s$,
\begin{equation}
f(s,\phi)=\sum_{m=0}^{\infty}e^{i m{\pi\over4}}A_m(\phi)s^m
\label{efeser}
\end{equation}
and insert this series in Eq.~(\ref{gedosult}). The values of $A_0(\phi)$ and
$A_2(\pi)$, which will be used below, are:
\begin{eqnarray}
A_0(\phi)&=&{1+\cos\phi\over \cos{\pi\over\alpha}-
\cos{\phi\over\alpha}},\nonumber\\
(2a)^{-{1\over2}}A_0(\phi)\Big|_{\phi=\pi^{\pm}}&=&\pm{i\alpha\over2\sin
{\pi\over\alpha}},\nonumber \\
A_2(\pi)&=&-{\cos{\pi\over\alpha}\over 2\sin^2{\pi\over\alpha}}.
\label{as}
\end{eqnarray}
Notice that the evaluation of $(2a)^{-1/2}A_0(\phi)$ when $\phi=\pm\pi$ is
actually a limit ($a=1+\cos\phi\approx 0$). It is
possible to show that all the
$A_{2m}(\phi)$ are finite at $\phi=\pm\pi$. The terms with odd $s$
in Eq.~(\ref{efeser}) cancel when
integrating, while the terms with even $s$ give after the substitution
$s=\tau\rho^{-1/2}$ a confluent hypergeometric function. In terms of
$S_m(x)$ functions, defined by Pauli \cite{pauli}, the propagator $G_2$ reads
\begin{eqnarray}
G_2(r,\theta;r',\theta';t)&=&-{m\sin{\pi\over\alpha}\over 4\pi^2 \hbar
t\alpha^2}
\exp\Big{\{}i
{m\over 2\hbar t\alpha^2}(r^2+r'^2)\Big{\}}e^{i(\rho+\pi/4)}\nonumber\\
& &\times\Big({2\over
a}\Big)^{1\over2}\sum_{m=0}^{\infty}i^m\Gamma\Big( m+{1\over2}\Big) A_{2m}
(\phi)S_m(a\rho)\rho^{-m}.
\label{gedosexp}
\end{eqnarray}
The behaviour of these $S_m(x)$ functions for large and small $x$ are
\begin{equation}
\begin{array}{lll}
S_m(x)&\approx -i x^{-{1\over2}}\left[1-\left(m+{1\over2}\right)(i x)^{-1}
+\cdots \right] ~, & \quad |x|>\!\!>1  \nonumber\\
S_m(x)&\approx \left(m-{1\over2}\right)^{-1}x^{1\over2} ~~, & \quad |x|\approx
0\hbox{~{\rm{and}}~} m>0  \nonumber\\
S_0(x)&\approx\pi^{1\over 2}e^{-i\pi/4} ~~, & \quad
|x|\approx 0 ~~.
\end{array}
\label{eses}
\end{equation}

\noindent We can now proceed with the analysis of the propagator $G_2$. There
are two interesting cases:

\subsection{$\rho(1+\cos\phi)\to\infty$}

This represents the large $mrr'/\hbar t\alpha^2$ limit, away from the
classical scattering
angles, {\it{i.e.}}, $\theta'-\theta\neq\pm\pi/\alpha$. Taking into account
Eqs.~(\ref{as}), (\ref{gedosexp}), (\ref{eses}) and going back to the original
variables $r, \theta$, the result is
\begin{eqnarray}
G_2(r,\theta;r',\theta';t)&=&{1\over 2\pi}\Big({m\over 2\pi i \hbar
t\alpha^2rr'}\Big)^{1\over2}{\sin{\pi\over\alpha}\over
\cos{\pi\over\alpha}-\cos(\theta'-\theta)} \nonumber\\
& &\times\exp\Big{\{}i
{m\over 2\hbar t\alpha^2}(r+r')^2\Big{\}} +{\cal{O}}(r^{-3/2}).
\label{gedosas}
\end{eqnarray}

\subsection{$\phi=\pi^{\pm}$}

These values of $\phi$ correspond to the classical scattering angles. The
parameter $-a=1+\cos\phi$ is now vanishing. That notwithstanding, the
singularity $1/\sqrt{a}$ in Eq.~(\ref{gedosexp}) is compensated by $A_0(\phi)$
if $m=0$ and by $S_m(a\rho)$ if $m>0$. This implies that
the asymptotic limit can be performed without finding any singularities at the
classical scattering angles, in contrast with the result of
applying the asymptotic limit directly to Eq.~(\ref{gedosdos}).

In this kinematic region we find a finite discontinuity:
\begin{eqnarray}
\lefteqn{G_2(r,\theta'-(\pi^{\pm}/\alpha);r',\theta';t)=}\nonumber\\
& &\pm{m\over 4\pi i \hbar t\alpha}\exp\Big{\{}i
{m\over 2\hbar t\alpha^2}(r+r')^2\Big{\}}\nonumber\\
& &+{i\over2\pi}\Big({i m\over 8\pi\hbar t\alpha^2
rr'}\Big)^{1\over2}\cot{\pi\over\alpha}\exp\Big{\{}i
{m\over 2\hbar t\alpha^2}(r+r')^2\Big{\}}\nonumber\\
& &+{\cal{O}}(r^{-3/2}).
\label{gedospi}
\end{eqnarray}
The two first terms in this expansion will be denoted by $G_{21}$
and $G_{22}$ respectively.
It should be noticed that the second term has the radial structure of a
scattered wave, and coincides with the result of taking the principal value in
the integral representation of $G_2$ given in Eq.~(\ref{geundos}). The
other terms, however, would be lost in so doing. In particular, the first
term represents a
discontinuous wave transmitted along the classical scattering angle, which
will be called ``subdominant'' because of its dependence on time. The wave
propagated by $G_{22}$ will be called ``leading''. The terms not included
in Eq.~(\ref{gedospi}) can be calculated by taking more
elements in the expansion (\ref{efeser}). These terms can be shown to be
continuous, and therefore do not contribute to the discontinuity of the
scattered wave at the classical scattering directions.

If $\phi=-\pi^{\pm}$ a similar analysis shows that the result is identical.
Therefore we shall not consider this case explicitly.

\section{Scattering of a wave packet: zero impact parameter}

\indent In this Section we consider the scattering of a Gaussian wave packet
by means of the propagator calculated in the previous Section. For the moment
we assume that the impact parameter is zero, and that the wave packet is
centered at $r\approx r_0$ and $\theta\approx\pi$. Its initial
momentum is $(k_0,0)$ in Cartesian coordinates; we will consider
that $r_0>\!\!>k_0^{-1}$:
\begin{equation}
\Psi_0(r',\theta',0)={1\over\sqrt{2\pi}\xi}\exp\Big{\{}i k_0
r'\cos\theta'-{1\over4\xi^2}\big(r'^2+r_0^2+2r'r_0\cos\theta'\big)\Big{\}}.
\label{pack}
\end{equation}
It is convenient to distinguish whether $\theta$ is
different from or equal to the classical scattering angle $\theta'\pm(\pi/
\alpha)$,
since in the first situation the relevant propagator is Eq.~(\ref{gedosas}),
whereas in the second one we need $G_1$ and Eq.~(\ref{gedospi}).

\subsection{$\theta\neq\theta'\pm(\pi/\alpha)$}

As stated before, the propagator is Eq.~(\ref{gedosas}), so that the
integration to be done is
\begin{eqnarray}
\Psi(r,\theta,t)&=&{1\over 2\pi}\Big({m\over 2\pi i \hbar
t\alpha^2r}\Big)^{1\over2}{\sin{\pi\over\alpha}\over
\cos{\pi\over\alpha}+\cos\theta}\int\limits_0^{2\pi}d\theta'\int
\limits_0^{\infty}
\sqrt{r'}dr'{1\over\sqrt{2\pi}\xi}\nonumber\\
& &\times\exp\Big{\{}i k_0
r'\cos\theta'-{1\over4\xi^2}\big(r'^2+r_0^2+2r'r_0\cos\theta'\big)\Big{\}}
\nonumber\\ & &
\times\exp\Big{\{}i{m\over 2\hbar t\alpha^2}(r+r')^2\Big{\}}.
\label{integ}
\end{eqnarray}
We have approximated $\theta'=\pi$ in the propagator but not in the initial
wave function. Following standard procedures we find that in the limit
$k_0>\!\!>r_0^{-1}$ the final wave function can be written as
\begin{equation}
\Psi(r,\theta,t)=\sqrt{{i\over r}}\,\,{1\over\sqrt{2\pi k_0}}\,\,
{\sin{\pi\over\alpha}\over
\cos{\pi\over\alpha}+\cos\theta}\,\,\Psi_{{\rm{free}}}(r,\alpha^2 t),
\label{psi}
\end{equation}
where $\Psi_{{\rm{free}}}$ denotes a freely propagating radial wave packet,
\begin{eqnarray}
\Psi_{{\rm{free}}}(r,\alpha^2t)&=&{1\over\sqrt{2\pi}\xi}\int
\limits_0^{\infty}dr'
\exp\Big{\{}-i k_0r'-{1\over4\xi^2}\big(r'-r_0\big)^2\Big{\}}\nonumber\\
& &\times\Big({m\over 2\pi i\hbar
t\alpha^2}\Big)^{1\over2}
\exp\Big{\{}i{m\over 2\hbar t\alpha^2}(r+r')^2\Big{\}}.
\label{free}
\end{eqnarray}
Note that the dependence on $t$ is through $\alpha^2 t$. This can be
interpreted as a time delay in the propagation of the scattered wave packet.
The delay $\Delta (t)$ of a scattered wave is usually due to
the dependence of the phase shifts on the energy,
as explained by Wigner's formula \cite{wigner} (see also \cite{gerjac}):
\begin{equation}
\Delta(t)=2{\partial\over\partial E}\delta_n(E)
\label{wigner}
\end{equation}
This cannot account for the time delay of $\Psi_{{\rm{free}}}$
because the partial wave analysis of this problem shows that the
phase shifts, Eq.~(\ref{phase}), do not depend on the
energy \cite{desjac}. We leave this question open for future clarification.

The scattering amplitude can be read from Eq.~(\ref{psi}), which
is the well-known result \cite{thooft}, \cite{desjac}.

\begin{equation}
f(k,\theta)={1\over\sqrt{2\pi k}}\,\,{\sin{\pi\over\alpha}\over
\cos{\pi\over\alpha}+\cos\theta},
\label{ampli}
\end{equation}

\subsection{$\theta\approx\theta'\pm(\pi/\alpha)$}

This angular range involves three main contributions: $G_1$ and the two terms
of $G_2$ shown in Eq.~(\ref{gedospi}). The contribution of $G_{22}$ to the
final wave function, denoted by $\Psi_{22}$ can be easily calculated:
\begin{equation}
\Psi_{22}(r,\theta'\pm{\pi\over\alpha},t)=-{1\over 2}
\sqrt{{i \over r}}\,\,
{1\over\sqrt{2\pi k_0}}\,\,\cot{\pi\over\alpha}
\,\,\Psi_{{\rm{free}}}(r,\alpha^2 t),
\label{psycho}
\end{equation}
Let us denote by $\Psi_{21}$ and $\Psi_1$ the contribution of $G_{21}$ and
$G_1$ to the final wave
function. $G_{21}$ presents a discontinuous behaviour in $\phi=\pi$ which
exactly compensates the discontinuity in $G_1$ due to the ``absorption''
of a new pole into the closed contour in Fig.~3. We shall show this
explicitly. Let $\delta$ be a small positive angle; the
discontinuity in $\Psi_{21}$ is
\begin{eqnarray}
& &
\Psi_{21}(r,\pi+{\pi\over\alpha}+\delta,t)-
\Psi_{21}(r,\pi+{\pi\over\alpha}-\delta,t)=\nonumber\\ & &
{m\over 2\pi i\hbar t\alpha}\int\limits_0^{2\pi}d\theta'\int\limits_0^{\infty}
r'dr'\,\,\Psi_0(r',\theta',0)\exp\Big{\{}{i m\over 2\pi\hbar t\alpha^2}
(r+r')^2\Big{\}} +{\cal{O}}(\delta),
\label{discon}
\end{eqnarray}
while the discontinuity in $\Psi_1$ is
\begin{eqnarray}
& &\Psi_1(r,\pi+{\pi\over\alpha}+\delta,t)-
\Psi_1(r,\pi+{\pi\over\alpha}-\delta,t)=\nonumber\\ & &
{m\over 2\pi i \hbar t\alpha}\int\limits_0^{2\pi}d\theta'
\int\limits_0^{\infty}
r'dr'\,\,\Psi_0(r',\theta',0)\exp\Big{\{}{i m\over 2\pi\hbar
t\alpha^2}(r^2+r'^2)
\Big{\}}\nonumber\\
& &\times\Big[\mathop{{\sum}'}_n \exp\Big{\{}{-i mrr'\over \hbar t\alpha^2}
\cos\Big(\pi+\alpha\delta-2\pi\alpha n\Big)\Big{\}} \nonumber\\
& &-\mathop{{\sum}'}_n \exp\Big{\{}{-i mrr'\over \hbar t\alpha^2}
\cos\Big(\pi-\alpha\delta-2\pi\alpha n\Big)\Big{\}}\Big].
\label{discodos}
\end{eqnarray}
Each sum includes all $n$ such that the argument of the cosine be in
$(-\pi,\pi)$. The range is different in each sum due to the presence of
$\delta$. More precisely, the maximum and minimum values of $n$ are
\begin{eqnarray}
n_{max}&=&\Big[{1\over\alpha}
\pm{\delta\over 2\pi}\Big]\approx\Big[{1\over\alpha}\Big] \nonumber\\
n_{min}&=&\Big[\pm{\delta\over 2\pi}\Big]+1=\left\{ \begin{array}
{ll}
1 & \mbox{if $+$}\\
0 & \mbox{if $-$}
\end{array}
\right.
\label{range}
\end{eqnarray}
Therefore, if we expand Eq.~(\ref{discodos}) in powers of $\delta$ all
leading terms cancel, except the one that comes from $n=0$ in the second
sum. The discontinuity in $\Psi_1$ is
\begin{eqnarray}
& &
\Psi_1(r,\pi+{\pi\over\alpha}+\delta,t)-
\Psi_1(r,\pi+{\pi\over\alpha}-\delta,t)=\nonumber\\ & &
-{m\over 2\pi i \hbar t\alpha}\int\limits_0^{2\pi}
d\theta'\int\limits_0^{\infty}
r'dr'\,\,\Psi_0(r',\theta',0)\exp\Big{\{}{i m\over 2\hbar
t\alpha^2}(r^2+r'^2+2rr'\cos(\alpha\delta)\Big{\}} \nonumber\\ & &\approx
-{m\over 2\pi i \hbar t\alpha}\int\limits_0^{2\pi}
d\theta'\int\limits_0^{\infty}
r'dr'\,\,\Psi_0(r',\theta',0)\exp\Big{\{}{i m\over 2\pi\hbar t\alpha^2}
(r+r')^2\Big{\}}
 +{\cal{O}}(\delta).
\label{discotres}
\end{eqnarray}

It is clear that the discontinuities in $\Psi_{21}$ and $\Psi_1$
cancelled out. Therefore we have shown that
the wave function is continuous along the classical scattering direction due
to the interference between the subdominant part of the scattered wave and the
transmitted wave. It can be shown that not only the discontinuities in the
scattered wave function, but also in its derivatives, are compensated by
those in the transmitted wave function. The leading part of the
scattered wave does not play any significant role in this
interference. This situation is reminiscent of Young's theory of
optical diffraction \cite{sommer}.

There is another relation which can be proven
within this framework: if we approach the classical scattering angle
$-\pi+\pi/\alpha$ from below we can write, in the limit of large $mrr'/\hbar
t\alpha^2$,
\begin{eqnarray}
\lefteqn{\Psi_{21}(r,-\pi+\pi/\alpha-\delta,t)=}\nonumber\\
& &-{m\over 4\pi i \hbar t\alpha}
\int\limits_0^{2\pi}
d\theta'\int\limits_0^{\infty}
r'dr'\,\,\Psi_0(r',\theta',0)\exp\Big{\{}{i m\over 2\pi\hbar t\alpha^2}
(r+r')^2\Big{\}}+{\cal{O}}(\delta),\nonumber\\
\lefteqn{\Psi_1(r,-\pi+\pi/\alpha-\delta,t)=}\nonumber\\
& &{m\over 2\pi i \hbar t\alpha}
\int\limits_0^{2\pi}
d\theta'\int\limits_0^{\infty}
r'dr'\,\,\Psi_0(r',\theta',0)\exp\Big{\{}{i m\over 2\pi\hbar t\alpha^2}
(r+r')^2\Big{\}}+{\cal{O}}(\delta).
\label{compar}
\end{eqnarray}
where in $\Psi_1$ only the $n=0$ term has been retained. The remaining terms
are negligible in the asymptotic limit. Also, $\Psi_{22}$ is much smaller than
$\Psi_{21}$ or $\Psi_1$ in that limit. Of course, $\delta$ is a
correspondingly small angle. Therefore we can conclude that
\begin{equation}
\Psi_{21}(r,\pi+\pi/\alpha-\delta,t)=-{1\over2}
\Psi_1(r,\pi+\pi/\alpha-\delta,t).
\label{conclu}
\end{equation}
If we denote the total asymptotic wave function in this angular region,
$\Psi_{21}+\Psi_1$, by $\Psi_{{\rm{total}}}$, we find
\begin{equation}
\Psi_{{\rm{total}}}(r,\pi+\pi/\alpha-\delta,t)={1\over2}
\Psi_1(r,\pi+\pi/\alpha-\delta,t).
\label{concludos}
\end{equation}
This corresponds to the verification done in \cite{pauli} of a general result
in the theory of diffraction, due to Sommerfeld \cite{sommer}: in the boundary
between shadow and light the total light amplitude is half the transmitted
amplitude.

There is a similar result for $\theta=\pi-\pi/\alpha$. The interpretation of
these equations is clear: the wave packet hits the scattering centre and
splits in two halves which propagate along the classical scattering
angles. This is analogous to the classical motion of a bunch of particles
approaching the scattering centre with zero average impact parameter.

\section{Scattering of a wave packet: non-zero impact parameter}

\indent In this last Section we generalize the previous results to
non-vanishing impact parameters. The initial Gaussian wave packet is now
centered at $(\rho,\theta_0)$ (polar coordinates); the impact parameter is
$b=\rho\sin\theta_0$. The momentum is the same as in Eq.~(\ref{pack}):
\begin{equation}
\Psi(r',\theta',0)={1\over\sqrt{2\pi}\xi}\exp\Big{\{}i k_0
r'\cos\theta'-{1\over4\xi^2}\big(r'^2+\rho^2-2r'\rho\cos(\theta'-\theta_0)
\big)\Big{\}}.
\label{packdos}
\end{equation}
The calculation follows the same steps as in the previous Section: if we
consider a scattering angle different from the classical one we must take
$G_{2}$ as the relevant propagator; otherwise we take $G_1$, $G_{21}$
and $G_{22}$.

Let us consider the first case. If the wave packet started its motion from
a long distance, the scattered wave can be written as
\begin{equation}
\Psi(r,\theta,t)=\sqrt{{i \over r}}\,\,{1\over\sqrt{2\pi k_0}}\,\,
{\sin{\pi\over\alpha}\over
\cos{\pi\over\alpha}+\cos\theta}\,\,e^{-{b^2\over4\xi^2}}\,\,
\Psi_{{\rm{free}}}(r,\alpha^2 t),
\label{psidos}
\end{equation}
Therefore, if $b>\!\!>\xi\/$ there is no significant
quantum scattering away from the
classical scattering angles. If $\theta$ is equal to these angles, the
relevant propagators are $G_1$, $G_{21}$ and $G_{22}$. The
contribution of $G_{22}$ is similar to Eq.~(\ref{psidos}) and hence can be
discarded, so that we are left with $G_1$ and $G_{21}$.

Let us consider that $\theta_0=\pi+\delta\/$
where $\delta\/$ is a small but finite
angle. When considering the wave packet in the remote past we will take
$\delta\to 0$ but it will never be exactly zero. This prevents the
contours in Fig.~3 from hitting the poles, and at the same time implies that
$G_{21}$ will not contribute. We recall here that this contribution to the
propagator arises as a discontinuity in the integral representation of $G_2$
which occurs only if the contour cannot be deformed to avoid the poles in
the real axis.

The contribution from $G_1$ depends on the sign of $\delta$. If $\delta>0$
the only contribution relevant in the asymptotic limit comes from $n=0$ and
$\theta=\pi+\pi/\alpha$ (other possibilities, like $n=1$ and
$\theta=-\pi+\pi/\alpha$ are physically equivalent). If $\delta<0$
we need to take $n=0$ and $\theta=\pi-\pi/\alpha$ instead, or any equivalent
choice. This can be written compactly in the notation
of Eq.~(\ref{concludos}):
\begin{eqnarray}
\delta>0\,\,&\Rightarrow&
\Psi_{{\rm{total}}}(r,\theta,t)=\Psi_1(r,-\pi+\pi/\alpha,t),\nonumber\\
\delta<0\,\,&\Rightarrow&
\Psi_{{\rm{total}}}(r,\theta,t)=\Psi_1(r,\pi-\pi/\alpha,t),
\label{finis}
\end{eqnarray}
This equations can be interpreted in the following way: the
wave packet follows the classical trajectory of a particle
with same initial position and velocity.

\section{Conclusions}
We can summarize our conclusions in four points:
\begin{enumerate}
\item The scattering amplitude coincides with the one found in \cite{thooft}
and \cite{desjac}.

\item The scattered wave packet is continuos everywhere. If the impact
parameter $b$ is not zero, it propagates like a classical particle. If $b=0$,
it hits the scattering centre and splits in two halves which propagate
along the classical scattering angles, plus a scattered
``spherical''  wave, thus confirming the qualitative analysis
in \cite{gerjac}.

\item If the impact parameter is zero, the continuity is due to
interference between the transmitted and the scattered parts of the wave
function along the classical scattering directions. This is similar to the
Aharonov-Bohm effect in the forward direction \cite{steli}; in both cases
the wave function undergoes a self-interference at the classical scattering
angles.

\item The time dependence of the scattered wave is modified due to the
presence of the massive scattering centre, see for example Eq.~(\ref{free}).
This calls for an explanation.

\end{enumerate}

\section{Appendix: time-independent scattering}

\indent In this Appendix we show that the same method can be applied to the
simpler case of time-independent scattering of plane waves in 2+1 dimensional
gravity. We find a similar cancellation of discontinuities along the
classical scattering angles but, being this a static problem, the cancelled
terms are not subdominant in time.

Let us recall the Deser-Jackiw solution for the time-independent scattering
problem \cite{desjac}:
\begin{eqnarray}
\Psi_{{\rm{in}}}(r,\theta)&=&\alpha\mathop{{\sum}'}_n e^{- kr\cos
\alpha(\theta-(2n+1)\pi)}, \nonumber \\
\Psi_{{\rm{sc}}}(r,\theta)&=&{1\over 4\pi}\int\limits_{-\infty}^{\infty}dy
\,e^{i kr\cosh y}\Big[\tan\Big( {i y\over 2\alpha}+
{\pi\over 2\alpha}+{\theta\over 2}\Big)-\tan\Big({i y\over 2\alpha}-
{\pi\over 2\alpha}+{\theta\over 2}\Big)\Big],
\label{geundosap}
\end{eqnarray}
where the primed sum includes only $n$ such that $\alpha(\theta-(2n+1)\pi)
\in (-\pi, \pi)$. The notation $\Psi_{{\rm{in}}}$ stands for the incoming
wave, and $\Psi_{{\rm{sc}}}$ for the scattered wave. We are going to
calculate the scattered wave
following the procedure described in \cite{pauli}.

To apply this method, we write $\Psi_{{\rm{sc}}}$ in an alternative way
by means of a trigonometric identity:
\begin{equation}
\Psi_{{\rm{sc}}}(r,\theta)={1\over 2\pi}\int
\limits_{-\infty}^{\infty} dy{\sin{\pi\over\alpha}\over
\cos{\pi\over\alpha}+\cos\Big({i y\over\alpha}+\theta\Big)}\,\,
e^{ikr\cosh y}.
\label{gedosdosap}
\end{equation}
As in the time-dependent case, if $\alpha^{-1}$
is an integer there is no scattered wave. If that
is not the case the integral can be performed in the
limit of large $kr$, where the leading contribution
comes from the small $y$ region. The result is
\begin{equation}
\Psi_{{\rm{sc}}}(r,\theta) \approx {\sin{\pi\over\alpha}\over 2\pi}
e^{ikr}\int\limits_{-\infty}
^{\infty}dy\,\, e^{{1\over 2}ikr y^2}
{1\over\cos{\pi\over\alpha}+\cos\theta-{i y\over\alpha}
\sin\theta+
{\cal{O}}(y^2)}.
\label{gedostresap}
\end{equation}
To proceed with the integration we will assume that
$\theta\neq\pi \pm \pi/\alpha\,\, {\rm{mod(2\pi)}}$,
{\it{i.e.}}, we keep away from the classical scattering
angles \cite{desjac}. In the large $kr$ limit, a Gaussian integration yields
\begin{equation}
\Psi_{{\rm{sc}}}(r,\theta)\approx \sqrt{{i \over r}}\,\,e^{i kr}
{1\over \sqrt{2\pi k}}\,\,{\sin{\pi\over\alpha}\over
\cos{\pi\over\alpha}+\cos\theta},
\label{gedosfinap}
\end{equation}
which gives the scattering amplitude found in \cite{thooft} and \cite{desjac}.
The behaviour of the scattered wave close to the
classical scattering angles can be determined as in the time-dependent
analysis; the analog of the change of variables (\ref{change}) in
Eq.~(\ref{gedosdosap}) is

\begin{eqnarray}
y&=&i \eta \nonumber \\
kr &=& \rho  \nonumber\\
\theta &=&-{\phi\over\alpha}+\pi.
\label{changeap}
\end{eqnarray}

In terms of the variables $\rho$ and $\phi$, the two physically
interesting situations are:

\subsection{$\rho(1+\cos\phi)\to\infty$}

This represents the large $kr$ limit, away from the classical
scattering angles, {\it{i.e.}}, $\theta\neq\pi\pm\pi/\alpha$. The analysis in
terms of $S_m$ functions of this kinematic region coincides with the
Gaussian integration of Eq.~(\ref{gedostresap}); the result is of
course Eq.~(\ref{gedosfinap}).

\subsection{$\phi=\pi^{\pm}$}

These values of $\phi$ correspond to the classical scattering angles.
In this case we find a discontinuous result. In the original
variables $r, \theta$, it reads
\begin{eqnarray}
\Psi_{{\rm{sc}}}(r,\pi-(\pi^{\pm}/\alpha))&=&
\pm{1\over 2}\alpha\,\,e^{i kr}
-\sqrt{{i\over 8\pi kr}}\,\,e^{i kr}\cot{\pi\over\alpha}\nonumber \\
& &+{\cal{O}}(r^{-3/2}).
\label{gedospiap}
\end{eqnarray}
The second term has the radial structure of a
scattered wave, and coincides with the result of taking the principal value in
the integral representation of $\Psi_{{\rm{sc}}}$ shown
in Eq.~(\ref{geundosap}). The first term represents a
discontinuous plane wave transmitted along the classical scattering angle,
whose discontinuity will be cancelled by another contribution coming from
$\Psi_{{\rm{in}}}$. The case $\phi=-\pi^{\pm}$ is no different.

The discontinuities cancel as in the time-dependent case. Let $\delta$ be a
small positive angle. The discontinuity in $\Psi_{sc}$ is
\begin{equation}
\Psi_{{\rm{sc}}}(r,\pi+{\pi\over\alpha}+\delta)-
\Psi_{{\rm{sc}}}(r,\pi+{\pi\over\alpha}-\delta)=\alpha\,\,e^{i kr}
+{\cal{O}}(\delta),
\label{disconap}
\end{equation}
while the discontinuity in $\Psi_{in}$ is
\begin{eqnarray}
& &\Psi_{{\rm{in}}}(r,\pi+{\pi\over\alpha}+\delta)-
\Psi_{{\rm{in}}}(r,\pi+{\pi\over\alpha}-\delta) \nonumber\\
& &=\alpha \Big[\mathop{{\sum}'}_n \exp\Big{\{}-i kr
\cos\Big(\pi+\alpha\delta-2\pi\alpha n\Big)\Big{\}} \nonumber\\
& &\quad -\mathop{{\sum}'}_n \exp\Big{\{}-i kr
\cos\Big(\pi-\alpha\delta-2\pi\alpha n\Big)\Big{\}}\Big].
\label{discodosap}
\end{eqnarray}
Each sum includes all $n$ such that the argument of its cosinus be in
$(-\pi,\pi)$. The range is different in each sum due to the presence of
$\delta$. More precisely, the maximum and minimum values of $n$ are
\begin{eqnarray}
n_{max}&=&\Big[{1\over\alpha}
\pm{\delta\over 2\pi}\Big]\approx\Big[{1\over\alpha}\Big] \nonumber\\
n_{min}&=&\Big[\pm{\delta\over 2\pi}\Big]+1=\left\{ \begin{array}
{ll}
1 & \mbox{if $+$}\\
0 & \mbox{if $-$}
\end{array}
\right.
\label{rangeap}
\end{eqnarray}
Therefore the only uncancelled leading term corresponds to $n=0$ in
the second sum. The discontinuity in $\Psi_{{\rm{in}}}$ is
\begin{equation}
\Psi_{{\rm{in}}}(r,\pi+{\pi\over\alpha}+\delta)-
\Psi_{{\rm{in}}}(r,\pi+{\pi\over\alpha}-\delta)=
-\alpha\,\,e^{i kr} +{\cal{O}}(\delta).
\label{discotresap}
\end{equation}
As expected, both discontinuities cancel. The wave function is
continuous at the classical scattering directions. It can be shown
that Sommerfeld's theorem holds also in this case, exactly as in the
time-dependent case.

\vspace{1cm}
{\bf{Acknowledgements:}}
The authors acknowledge Professor Roman Jackiw for suggesting this problem
and for many helpful comments, and Professor Stanley Deser for a careful
reading of the manuscript. We are indebted to Prof. Negele and the
CTP for hospitality.

This work vas supported in part by the Spanish CICYT (M.A.), by the
Brazilian CAPES (F.M.C.F.) and by Padua University and Aldo Gini
Foundation (L.G.)



\end{document}